\documentstyle[12pt]{article}
\catcode`\@=11
\def\theequation{\arabic{equation}}
\def\section{\@startsection{section}{1}{\z@}{3.5ex plus 1ex minus
   .2ex}{2.3ex plus .2ex}{\large\bf}}
\def\thesection{\arabic{section}.}

\def\appendix{\setcounter{section}{0}
        \def\thesection{Appendix.} }
\def\eqnarray{\let\@currentlabel=\theequation\refstepcounter{equation}
    \global\@eqnswtrue
    \global\@eqcnt\z@\tabskip\@centering\let\\=\@eqncr
    $$\halign to \displaywidth\bgroup\@eqnsel\hskip\@centering
      $\displaystyle\tabskip\z@{##}$&\global\@eqcnt\@ne
       \hfil${{}##{}}$\hfil
      &\global\@eqcnt\tw@ $\displaystyle\tabskip\z@{##}$\hfil
       \tabskip\@centering&\llap{##}\tabskip\z@\cr}
\def\lefteqn#1{\hbox to 4\arraycolsep{$\displaystyle #1$\hss}}

\long\def\@makefntext#1{\parindent 0cm\noindent
\hbox to 1em{\hss$^{\@thefnmark}$}#1}
\def\IR{{\hbox{{\rm I}\kern-.2em\hbox{\rm R}}}}
\def\IH{{\hbox{{\rm I}\kern-.2em\hbox{\rm H}}}}
\def\IC{{\ \hbox{{\rm I}\kern-.6em\hbox{\bf C}}}}
\def\IZ{{\hbox{{\rm Z}\kern-.4em\hbox{\rm Z}}}}
\def\rref#1{(\ref{#1})}
\def\t{\tau}
\newcommand{\beq}{\begin{equation}}
\newcommand{\eeq}{\end{equation}}
\newcommand{\ddet}{\det\hspace{-2pt}}
\newcommand{\NPB}[1]{{\sl Nucl.~Phys.}~{\bf B#1}}
\newcommand{\Ann}[1]{{\sl Ann.~Phys.}~{\bf #1}}

\newcommand{\PLB}[1]{{\sl Phys.~Lett.}~{\bf B#1}}

\newcommand{\PRL}[1]{{\sl Phys.~Rev.~Lett.}~{\bf #1}}

\newcommand{\MPLA}[1]{{\sl Mod.~Phys.~Lett.}~{\bf A#1}}
\newcommand{\IJMPA}[1]{{\sl Int.~J.~Mod.~Phys.}~{\bf A#1}}

\newcommand{\CQG}[1]{{\sl Class.~Quant.~Grav.}~{\bf #1}}
\newcommand{\PRD}[1]{{\sl Phys.~Rev.}~{\bf D#1}}

\newcommand{\JMP}[1]{{\sl J.~Math.~Phys.}~{\bf #1}}
\begin{document}
%
%
%
%
\def\citen#1{%
\edef\@tempa{\@ignspaftercomma,#1, \@end, }
\edef\@tempa{\expandafter\@ignendcommas\@tempa\@end}%
\if@filesw \immediate \write \@auxout {\string \citation {\@tempa}}\fi
\@tempcntb\m@ne \let\@h@ld\relax \let\@citea\@empty
\@for \@citeb:=\@tempa\do {\@cmpresscites}%
\@h@ld}
%
\def\@ignspaftercomma#1, {\ifx\@end#1\@empty\else
   #1,\expandafter\@ignspaftercomma\fi}
\def\@ignendcommas,#1,\@end{#1}
%
%
\def\@cmpresscites{%
 \expandafter\let \expandafter\@B@citeB \csname b@\@citeb \endcsname
 \ifx\@B@citeB\relax 
    \@h@ld\@citea\@tempcntb\m@ne{\bf ?}%
    \@warning {Citation `\@citeb ' on page \thepage \space undefined}%
 \else
    \@tempcnta\@tempcntb \advance\@tempcnta\@ne
    \setbox\z@\hbox\bgroup 
    \ifnum\z@<0\@B@citeB \relax
       \egroup \@tempcntb\@B@citeB \relax
       \else \egroup \@tempcntb\m@ne \fi
    \ifnum\@tempcnta=\@tempcntb 
       \ifx\@h@ld\relax 
          \edef \@h@ld{\@citea\@B@citeB}%
       \else 
          \edef\@h@ld{\hbox{--}\penalty\@highpenalty \@B@citeB}%
       \fi
    \else   
       \@h@ld \@citea \@B@citeB \let\@h@ld\relax
 \fi\fi%
 \let\@citea\@citepunct
}
%
\def\@citepunct{,\penalty\@highpenalty\hskip.13em plus.1em minus.1em}%
%
%
\def\@citex[#1]#2{\@cite{\citen{#2}}{#1}}%
%
%
\def\@cite#1#2{\leavevmode\unskip
  \ifnum\lastpenalty=\z@ \penalty\@highpenalty \fi 
  \ [{\multiply\@highpenalty 3 #1
      \if@tempswa,\penalty\@highpenalty\ #2\fi 
    }]\spacefactor\@m}
\let\nocitecount\relax  
%
\begin{titlepage}
\vspace{.5in}
\begin{flushright}
UCD-95-10\\
gr-qc/9504033\\
April 1995\\
(revised June 1995)\\
\end{flushright}
\vspace{.5in}
\begin{center}
{\Large\bf
A Phase Space Path Integral\\[.5ex] for (2+1)-Dimensional Gravity}\\
\vspace{.4in}
{S.~C{\sc arlip}\footnote{\it email: carlip@dirac.ucdavis.edu}\\
       {\small\it Department of Physics}\\
       {\small\it University of California}\\
       {\small\it Davis, CA 95616}\\{\small\it USA}}
\end{center}

\vspace{.5in}
\begin{center}
{\large\bf Abstract}
\end{center}
\begin{center}
\begin{minipage}{4.5in}
{\small I investigate the relationship between the phase space path
integral in (2+1)-dimensional gravity and the canonical quantization
of the corresponding reduced phase space in the York time slicing.
I demonstrate the equivalence of these two approaches, and discuss
some subtleties in the definition of the path integral necessary
to prove this equivalence.
}
\end{minipage}
\end{center}
\end{titlepage}
\addtocounter{footnote}{-1}

Over the past several years, (2+1)-dimensional general relativity has
become a popular model in which to explore the conceptual foundations
of quantum gravity \cite{Korea}.  But although a few papers have been
written about path integral methods \cite{Mart,Witb,Gonzalez,Mazur,
Amano,CarCos,Hosoya,Cartop}, the primary focus of research has been on
various forms of canonical quantization.  The purpose of this paper is
to briefly describe a phase space path integral, both to display its
equivalence to canonical quantization on a reduced phase space and to
underline the assumptions needed to demonstrate this equivalence.

In one sense, the results of this paper are obvious: there exist
fairly general formal proofs of the equivalence of phase space path
integration and reduced phase quantization \cite{Henna,HennTeit,Kun}.
Indeed, the path integral for general relativity may be {\em defined\/}
by the requirement that it give the correct reduction to the physical
phase space \cite{Faddeev,Fradkin,FradVil}.  But these proofs involve
subtle assumptions about measures, gauge-fixing procedures, and ranges
of integration \cite{Gov}, and they are particularly tricky when one
deals with general relativity, a theory in which the Hamiltonian
constraint plays a rather peculiar role \cite{Guven}.  It is therefore
useful to explore a simple model in which all of the details can be
made explicit.

In this paper I concentrate on (2+1)-dimensional gravity on a manifold
with the topology $\IR\!\times\!\Sigma$, where $\Sigma$ is a closed
orientable
surface of genus $g\!>\!0$.  The Hamiltonian reduction of this model in
the York time slicing has been completely analyzed by Moncrief \cite{Mon,
HosNak}.  Thanks to the finite number of physical degrees of freedom of
(2+1)-dimensional general relativity, a simple and complete description
of the reduced phase space is possible.  It should be apparent, however,
that much of the analysis presented here also applies, at least formally,
to 3+1 dimensions as well.

The canonical action for (2+1)-dimensional gravity is\footnote{Here,
$g_{ij}$ and $R$ refer to the induced metric and scalar curvature of
a time slice, while the spacetime metric and curvature are denoted
${}^{\scriptscriptstyle(3)}\!g_{\mu\nu}$ and ${}^{\scriptscriptstyle(3)}
\!R$.  Roman indices $i,j,\dots$ are spatial indices, raised and lowered
with the spatial metric $g_{ij}$; Greek indices $\mu,\nu,\dots$ are
spacetime indices.}
\beq
I_{\hbox{\scriptsize grav}}
  = \int\!d^3x \sqrt{-{}^{\scriptscriptstyle(3)}\!g}\>
  {}^{\scriptscriptstyle(3)}\!R
  = \int dt\int\nolimits_\Sigma d^2x \bigl(\pi^{ij}{\dot g}_{ij}
               - N^i{\cal H}_i -N{\cal H}\bigr) ,
\label{a1}
\eeq
where the momentum and Hamiltonian constraints take the form
\begin{eqnarray}
{\cal H}_i &=& -2\nabla_j\pi^j_{\ i} \nonumber\\
{\cal H} &=&
{1\over\sqrt{g}}\,g_{ij}g_{kl}(\pi^{ik}\pi^{jl}-\pi^{ij}\pi^{kl})
  -\sqrt{g}\,R .
\label{a2}
\end{eqnarray}
The phase space path integral is
\beq
Z = \int [d\pi^{ij}][dg_{ij}][dN^i][dN]
    \exp\left\{ i I_{\hbox{\scriptsize grav}}[\pi,g] \right\} ,
\label{a3}
\eeq
but the first class constraints ${\cal H}_\mu = ({\cal H},{\cal H}_i)$
generate a set of transformations that must be gauge fixed in order for
this expression to be well-defined.  For gauge conditions $\chi^\mu=0$,
the path integral becomes \cite{HennTeit,Faddeev}
\beq
Z=\int [d\pi^{ij}][dg_{ij}][dN^i][dN]
  \delta[\chi^\mu]\ddet|\{{\cal H}_\mu,\chi^\nu\}|
  \exp\left\{ i I_{\hbox{\scriptsize grav}}[\pi,g] \right\} ,
\label{a4}
\eeq
where $\{\ ,\, \}$ denotes the Poisson bracket.  Our goal is to evaluate
this integral, reducing it to a quantum mechanical path integral over
the finitely many physical degrees of freedom of (2+1)-dimensional
gravity.

It is useful to start with a decomposition of the fields $g_{ij}$
and $\pi^{ij}$.  Let us assume for now that $\Sigma$ has genus $g\!>\!1$;
the case of the torus ($g\!=\!1$) will be discussed briefly below.
A standard result of Riemann surface theory, the uniformization theorem,
implies that any metric on $\Sigma$ can then be written in the form
\cite{Abikoff,Farkas,Alvarez}
\beq
g_{ij} = f^* e^{2\lambda}{\bar g}_{ij}(\t_r) ,
\label{a5}
\eeq
where $f$ is a diffeomorphism (typically generated by a vector field
$\xi^i$), $\lambda$ is a conformal factor, and the ${\bar g}_{ij}(\t_r)$
are a finite-dimensional family of fixed reference metrics of constant
curvature $k\!=\!-1$, parametrized by $6g-6$ moduli $\t_r$.
A related parametrization can be given for the momenta $\pi^{ij}$:
\beq
\pi^{ij} = {1\over2}g^{ij}\pi + \sqrt{g}(PY)^{ij}
  + \sqrt{g}\, p_r\Psi^{(r)\,ij} ,
\label{a8}
\eeq
where $P$ is an operator taking vectors $\xi^i$ to traceless tensors
$h_{ij}$,
\beq
(P\xi)_{ij} = \nabla_i\xi_j + \nabla_j\xi_i - g_{ij}\nabla_k\xi^k, \qquad
(P^\dagger h)_i = -2\nabla^jh_{ij} ,
\label{a7}
\eeq
and the $\Psi^{(r)}$ are a basis of $\hbox{ker}P^\dagger$, that is,
transverse traceless tensors, or in the language of Riemann surfaces,
quadratic differentials.  Note that the dimension of $\hbox{ker}
P^\dagger$ is $6g-6$, the same as the dimension of the moduli space;
indeed, the $\Psi^{(r)}$ are a basis for the tangent space of the moduli
space of $\Sigma$ \cite{Mon,Alvarez}.

In terms of these new variables, the constraints \rref{a2} become
\begin{eqnarray}
{\cal H}_i &=& \sqrt{g}(P^\dagger PY)_i - \nabla_i\pi  \nonumber\\
{\cal H} &=&
-{1\over2}{\pi^2\over\sqrt{\bar g}}e^{-2\lambda}
+\sqrt{\bar g}e^{-2\lambda}p_rp_s
   {\bar g}^{ij}{\bar g}^{kl}\Psi^{(r)}_{ik}\Psi^{(s)}_{jl} \\
&\phantom{=}& \phantom{-{1\over2}{\pi^2\over\sqrt{\bar g}}e^{-2\lambda}+}
+\sqrt{\bar g}e^{-2\lambda}
   {\bar g}^{ij}{\bar g}^{kl}(PY)_{ik}(PY)_{jl}
+\sqrt{\bar g}(2{\bar\Delta}\lambda - k) ,\nonumber
\label{a9}
\end{eqnarray}
where $\bar\Delta$ is the Laplacian with respect to $\bar g_{ij}$.
We would like to change variables in the path integral from $(g_{ij},
\pi^{ij})$ to $(\t_r,\lambda,f,p_s,\pi,Y^i)$.  To find the Jacobian
for this transformation, we follow the approach introduced in string
theory by Alvarez \cite{Alvarez,MN}.  (For an application to the
covariant path integral in quantum gravity, see \cite{Bern}.)  We
start by defining inner products
\begin{eqnarray}
\langle \delta g, \delta g \rangle
 &=&\int_\Sigma d^2x \sqrt{g} g^{ij}g^{kl}\delta g_{ik}\delta g_{jl}
\nonumber\\
\langle \delta\xi, \delta\xi \rangle
 &=&\int_\Sigma d^2x \sqrt{g} g^{ij}\delta\xi_i\delta\xi_j
\\
\langle \delta \lambda, \delta \lambda \rangle
 &=&\int_\Sigma d^2x \sqrt{g}\, (\delta \lambda)^2
\nonumber
\label{a10}
\end{eqnarray}
on the tangent space to the space of metrics on $\Sigma$.  Corresponding
to the parametrization \rref{a5}, an arbitrary infinitesimal deformation of
the metric $g_{ij}$ admits an orthogonal decomposition\footnote{$\delta
\tilde\lambda$ and $\delta\tilde\xi$ are infinitesimal Weyl transformations
and diffeomorphisms, up to linear shifts.} \cite{Alvarez,MN}
\beq
\delta g_{ij}
 = 2(\delta\tilde\lambda) g_{ij}+(P(\delta\tilde\xi))_{ij}
 + \delta\t_r\, T^{rs}\Psi^{(s)}_{ij} ,
\label{a6}
\eeq
where the last term is the orthogonal projection of the modular deformation
of $g_{ij}$ onto $\hbox{ker}P^\dagger$, that is,
\beq
T^{rs} = \langle\chi^{(r)},\Psi^{(u)}\rangle
  \langle\Psi^{(u)},\Psi^{(s)}\rangle^{-1} , \quad\hbox{with} \quad
\chi^{(r)}_{ij} = {\partial g_{ij}\over\partial\t_r} .
\label{a6a}
\eeq
Now consider the simple Gaussian path integral
\begin{eqnarray}
1 &=& \int[dg_{ij}]\, e^{i \langle\delta g,\delta g\rangle} \\
  &=& \int d^n(\delta\t)\,[d(\delta\tilde\lambda)][d(\delta\tilde\xi)]J\,
  e^{i \langle \delta\tilde\xi, P^\dagger P\delta\tilde\xi\rangle}
  e^{8i \langle\delta\tilde\lambda,\delta\tilde\lambda\rangle}
  e^{i\delta\t_r\delta\t_sT^{ru}T^{sv}\langle\Psi^{(u)},\Psi^{(v)}\rangle}
. \nonumber
\label{a11}
\end{eqnarray}
Evaluating the integrals over $\delta\t$, $\delta\tilde\lambda$, and
$\delta\tilde\xi$, we easily find that the Jacobian $J$ is
$$
J_g = \ddet|P^\dagger P|^{1/2} \ddet|T| \,
  \ddet|\langle\Psi^{(u)},\Psi^{(v)}\rangle|^{1/2}.
$$
This Jacobian was derived by
considering integrals on the tangent space to the space of metrics, but
a simple argument shows that it is equal to the Jacobian for the integral
over the $g_{ij}$ \cite{MN}.  The $\pi^{ij}$ integral gives a similar
Jacobian,
$$
J_{\pi} = \ddet|P^\dagger P|^{1/2}
  \ddet|\langle\Psi^{(u)},\Psi^{(v)}\rangle|^{1/2} ,
$$
which combines with $J_g$ to give a total Jacobian
\beq
J = \ddet|P^\dagger P|\,\ddet|T|\,\ddet|\langle\Psi^{(u)},\Psi^{(v)}\rangle|
  = \ddet|P^\dagger P|\,\ddet|\langle\chi^{(u)},\Psi^{(v)}\rangle| .
\label{a13a}
\eeq

If we now change variables and integrate over $N^\mu$, the path integral
\rref{a4} becomes
\begin{eqnarray}
Z=\int d^n\!p\,&d&^n\!\t\ddet|\langle\chi^{(u)},\Psi^{(v)}\rangle| \\
  \times\!\int[&d&(\pi/\!\sqrt{g})][d\lambda][dY][d\xi]
  \ddet|P^\dagger P|\, \delta[\chi^\mu]\,
  \delta[{\cal H}_\nu/\!\sqrt{g}]\ddet|\{{\cal H}_\mu,\chi^\nu\}|
  e^{iI_{\hbox{\scriptsize grav}}[p,\t,\lambda,\pi]} . \nonumber
\label{a14}
\end{eqnarray}
The factors of $1/\!\sqrt{g}$ in the delta functionals are somewhat
conventional, but can be viewed as coming from the inner products
\rref{a10} and the rule
$$
\int [da] e^{i\langle a,b \rangle} = \delta[b] .
$$
The integral over $Y_i$ is now straightforward; from \rref{a9},
$$\int [dY_i] \ddet|P^\dagger P|\, \delta[{\cal H}_i/\!\sqrt{g}] = 1 ,$$
where from now on we set
$Y_i = (P^\dagger P)^{-1}\nabla_i{\pi/\!\sqrt{g}} .$

To proceed further, we must make a partial choice of gauge fixing.  The
momentum constraints ${\cal H}_i$ generate ordinary spatial diffeomorphisms,
and their treatment is straightforward.  The Hamiltonian constraint $\cal
H$, on the other hand, does {\em not\/} generate time reparametrizations,
as one might naively expect, although the corresponding transformations
are related on shell \cite{Guven}.  Nevertheless, $\cal H$ generates an
invariance that must be gauge fixed \cite{Woodard}.  Following Moncrief,
let us do so by choosing the York time slicing \cite{York}
\beq
\chi = \pi/\!\sqrt{g} - T = 0 ,
\label{a15}
\eeq
where $T$ is a time coordinate.  Observe that with this gauge choice,
$\nabla_i\pi=0$, and hence $Y_i=0$.  Note  also that $\chi$ is a spatial
scalar, so $\{{\cal H}_i,\chi\} = 0$ when $\chi=0$.  The determinant
$\ddet|\{{\cal H}_\mu,\chi^\nu\}|$ therefore splits into a product
$$\ddet|\{{\cal H},\chi\}|\,\ddet|\{{\cal H}_i,\chi^j\}| .$$

The first term is easily evaluated, using the canonical commutators
\beq
\{ g_{ij}(x),\pi^{kl}(x')\} = \delta^{kl}_{ij}\delta^2(x-x') ,
\label{a16}
\eeq
and yields
\beq
\ddet|\{{\cal H},\chi\}|
=\det\Bigl| -\Delta + {1\over g}\pi_{ij}\pi^{ij} \Bigr|
= \det\Bigl|e^{-2\lambda}\Bigl( -\bar\Delta + {T^2\over2}e^{2\lambda}
  + e^{-2\lambda}p_rp_s{\bar g}^{ij}{\bar g}^{kl}
  \Psi^{(r)}_{ik}\Psi^{(s)}_{jl}\Bigr)\Bigr|
\label{a17}
\eeq
at $Y_i=0$ and $\pi/\!\sqrt{g}=T$.
We could next choose a gauge condition $\chi^i=0$ and evaluate the
remaining determinant, but we need not do so.  Almost everything
in the path integral is invariant under spatial coordinate
transformations; the only remaining $\xi$-dependent terms are
$$\int[d\xi]\delta[\chi^i]\ddet|\{{\cal H}_i,\chi^j\}| = 1 .$$
The path integral \rref{a14} thus simplifies to
\beq
Z=\int d^n\!p\, d^n\!\t \ddet|\langle\chi^{(u)},\Psi^{(v)}\rangle|
  \int[d\lambda]\delta[{\cal H}/\!\sqrt{g}]\ddet|\{{\cal H},\chi\}|\,
  e^{iI_{\hbox{\scriptsize grav}}[p,\t,\lambda]}
\label{a18}
\eeq
with $\ddet|\{{\cal H},\chi\}|$ given by \rref{a17}.

We can now use the remaining delta functional to evaluate the integral
over $\lambda$.  From eqn.\ \rref{a9},
\beq
{\cal H}/\!\sqrt{g} = 2e^{-2\lambda} \left(
{\bar\Delta}\lambda - {k\over2} - {T^2\over4}e^{2\lambda}
+ {1\over2} e^{-2\lambda}p_rp_s
   {\bar g}^{ij}{\bar g}^{kl}\Psi^{(r)}_{ik}\Psi^{(s)}_{jl} \right) .
\label{a19}
\eeq
Moncrief has shown that the equation ${\cal H}=0$ has a unique
solution $\lambda = {\bar\lambda}(p,\t,T)$ \cite{Mon}.  Hence
$$ \delta[{\cal H}/\!\sqrt{g}]
 = \det\Biggl|{\delta\ \over\delta\lambda}({\cal H}/\!\sqrt{g})\Biggr|^{-1}
 \delta[\lambda - \bar\lambda] ,$$
and the determinant is easily seen to exactly cancel \rref{a17} when
${\cal H}\!=\!0$.  We thus obtain
\beq
Z=\int d^n\!p\, d^n\!\t \ddet|\langle\chi^{(u)},\Psi^{(v)}\rangle|
  \, e^{i\bar I_{\hbox{\scriptsize grav}}[p,\t]} ,
\label{a20}
\eeq
where $\bar I$ is the action evaluated at the solution of the
constraints.

Finally, let us consider the remaining determinant $\ddet|\langle
\chi^{(u)},\Psi^{(v)}\rangle|$.  I have defined the reduced phase
space coordinate $p_r$ by the decomposition \rref{a8}.  Moncrief, on
the other hand, uses a slightly different parametrization,
\beq
\tilde p_r = \int_\Sigma d^2x\, e^{2\lambda}(\pi^{ij}-{1\over2}g^{ij}\pi)
  {\partial {\bar g}_{ij}\over\partial\t_r}
  = \langle p_s\Psi^{(s)},\chi^{(r)}\rangle
  = \langle\chi^{(r)},\Psi^{(s)}\rangle p_s .
\label{22}
\eeq
By changing variables to $\tilde p_r$, we can write the path integral
\rref{a20} as
\beq
Z=\int d^n\!\tilde p\, d^n\!\t \,
  e^{i\bar I_{\hbox{\scriptsize grav}}[\tilde p,\t]} .
\label{a23}
\eeq
An explicit expression for the action $\bar I$ may be obtaind by inserting
eqns.\ \rref{a5} and \rref{a8} into the term $\pi^{ij} \dot g_{ij}$ in
eqn.\ \rref{a1}; a simple computation gives
\beq
\bar I_{\hbox{\scriptsize grav}}[\tilde p,\t]
  = \int dT \left( \tilde p_r{d\t_r\over dT} - H(\tilde p,\t,T) \right),
  \qquad
H = \int_\Sigma d^2x\, \sqrt{\bar g} e^{2\bar\lambda(\tilde p,\t,T)} .
\label{a21}
\eeq
This is precisely Moncrief's reduced phase space action \cite{Mon}, and
eqn.\ \rref{a23} is exactly the right quantum mechanical path integral
for the corresponding reduced phase space quantum theory.  To the extent
that an ordinary quantum mechanical path integral for a system with
finitely many degrees of freedom is equivalent to canonical quantization,
we have therefore reproduced the quantum theory described in reference
\cite{Korea}.

A similar analysis is possible when $\Sigma$ is a torus, although a few
complications occur.  Decompositions of the form \rref{a5} and \rref{a8}
again exist; the $\bar g_{ij}$ are now a two-parameter family of metrics
of curvature $k\!=\!0$, normalized to unit volume.  The operator $P^\dagger
P$ now has zero modes, however, corresponding to conformal Killing vectors;
its determinant is thus identically zero.  A similar phenomenon occurs in
string theory, where a careful analysis shows that it causes no problems
(see \cite{MN}), essentially because the zero modes should be omitted from
the decompositions \rref{a8} and \rref{a6}.  A further complication arises
from the existence of classical configurations with $p_r\!=\!0$, for which
the constraint ${\cal H}\!=\!0$ has solutions only when $T$ vanishes.  For
these configuations, the gauge choice \rref{a15} is no longer admissible.
Such exceptional solutions also complicate the analysis in the Chern-Simons
formulation \cite{MarLou}, but they form a set of measure zero in the
space of solutions, and can probably be safely omitted from the path
integral.  If we do so, we again recover expressions of the form \rref{a23}
and \rref{a21}, and the path integral again reproduces canonical
quantization.

This result is not surprising, but it is worthwhile to review the
assumptions needed to reach this conclusion.  Four in particular are
important:
\setlength{\leftmargini}{1.7em}
\begin{enumerate}

\item {\bf Gauge fixing:} The gauge choice \rref{a15} is the phase
space version of the York time slicing condition, in which the mean
(extrinsic) curvature $\hbox{Tr}K\!=\!T$ is used as a time coordinate.
Moncrief has shown that for {\em solutions\/} of the field equations
in 2+1 dimensions, this is a good global coordinate choice, at least
in the domain of dependence of an initial spacelike surface $\Sigma$
of genus $g\!>\!1$.  But the path integral involves a sum over all
spacetimes, including geometries for which $\hbox{Tr}K\!=\!T$ is
certainly not a good time slicing.  This is not necessarily a
contradiction, since the relationship between $\pi$ and $\hbox{Tr}K$
also breaks down off shell, but there is clearly more to be understood
here.  A useful starting point might be a careful analysis of the
exceptional $p_r\!=\!0$ solutions for genus $1$, which are classical
solutions for which the gauge choice \rref{a15} fails.

\item {\bf Conformal anomalies:} My derivation of the reduced phase
space path integral \rref{a23} involved no determinants of the form
$\ddet|e^{2\lambda}|$.  Depending on the precise choice of the measure,
however, such terms could appear.  For example, it is not obvious a
priori whether the integral over the lapse function should give
$\delta[{\cal H}/\!\sqrt{g}]$ or, say, $\delta[{\cal H}]$; these two
possibilities differ by such a determinant.

Similar determinants appear as conformal anomalies in noncritical
string theory, where it has been argued that they lead to a Liouville
action for $\lambda$ \cite{DHoker},
$$
I = \int_\Sigma d^2x\, \sqrt{\bar g}\left( {1\over2}
  \lambda\bar\Delta\lambda + k\lambda + \mu_0^{\,2}e^{2\lambda}\right) .
$$
Since the value of $\lambda$ is fixed by a delta functional in our path
integral, such a term would not lead to new degrees of freedom, but it
would change the action \rref{a21}.  For a torus universe, $k\!=\!0$
and the Hamiltonian constraint requires that $\lambda$ be constant,
so the only effect of a Liouville term would be to multiplicatively
renormalize the reduced phase space Hamiltonian $H$.  For genus $g\!>\!1$,
however, the effect would be more significant.

\item {\bf Lapse integration:} A crucial element of this derivation
was the appearance of a delta functional $\delta[{\cal H}/\!\sqrt{g}]$
coming from the integration over the lapse $N$.  Such a term requires
an integration range from $-\infty$ to $\infty$ and a Lorentzian
signature ($iN{\cal H}$ rather than $N{\cal H}$ in the exponent).  As
Teitelboim has noted \cite{Teit2}, this is not a unique choice: for
instance, one could instead define a ``causal'' amplitude by integrating
over only positive values of $N$.  It should be clear that such an
amplitude is {\em not\/} equivalent to that coming from reduced phase
space quantization.

\item {\bf Choices of time slicing:} This paper has considered only
one special choice of time slicing.  In canonical quantization, it
is not at all clear that different choices lead to equivalent quantum
theories \cite{Kuchar}.  In the path integral formalism, on the other
hand, a general theorem due to Fradkin and Vilkovisky states that the
path integral is independent of the gauge-fixing function $\chi$
\cite{HennTeit,FradVil}.  It would be interesting to try to apply this
theorem to a concrete example in 2+1 dimensions, and to work out
implications for canonical quantization.  In particular, one might
compare this slicing with Teitelboim's ``proper time gauge'' $\partial_t
N\!=\!0$ \cite{Teit3}.  For the torus universe, this gauge is classically
equivalent to the York slicing, but the two need not be equivalent off
shell, so a comparison could be illuminating.

\end{enumerate}

Finally, let me briefly discuss the extension of this analysis to
3+1 dimensions.  As in 2+1 dimensions, the spatial metric admits a
decomposition of the form \rref{a5}, where the metrics ${\bar g}_{ij}$
can be determined by the Yamabe condition $R(\bar g)\!=\!\hbox{\em const.
\/}$ along with some suitable spatial gauge condition \cite{Bern,Schoen}.
In contrast to the (2+1)-dimensional model, the ${\bar g}_{ij}$ now
form an infinite-dimensional space, whose properties are not fully
understood.  Nevertheless, one can at least formally extend many of the
results of this paper.  The decompositions analogous to \rref{a8} and
\rref{a6} are discussed in reference \cite{Yorkb}, and the corresponding
Jacobians have been derived, albeit in a rather different context, in
references \cite{Mazur} and \cite{Bern}. It is not hard to show that
the Jacobian analogous to $\ddet|P^\dagger P|$ is again cancelled by
terms coming from the integration over $Y$, and that in the York time
slicing \rref{a15}, the determinant $\ddet|\{{\cal H},\chi\}|$ is
cancelled, up to possible conformal determinants, when one integrates
the delta functional $\delta[{\cal H}]$.  What is missing, however, is
an explicit description of the resulting (infinite dimensional) reduced
phase space and the measure corresponding to that of \rref{a23}, i.e.,
a three-dimensional version of moduli space.

\vspace{1ex}
\begin{flushleft}
\large\bf Acknowledgements
\end{flushleft}

This work was supported in part by National Science Foundation grant
PHY-93-57203 and Department of Energy grant DE-FG03-91ER40674.

\end{document}